\RequirePackage{fix-cm}
\documentclass[smallextended]{svjour3}       % onecolumn (second format)
\smartqed  % flush right qed marks, e.g. at end of proof
\usepackage{amssymb}
\usepackage{graphicx}
\usepackage{subfigure}
\usepackage{xcolor}
\usepackage{bm}

\def\oneht{\textstyle{1\over 2} }
\def\si{^1 \hskip -0.03in S _0}
\def\siii{^3 \hskip -0.025in S _1}
\newcommand{\nn}{\nonumber}
\newcommand{\LOss}{\scriptscriptstyle{LO}}
\newcommand{\NLOss}{\scriptscriptstyle{NLO}}

\def\llra{{\relbar\joinrel\longrightarrow}}

\def\mapup#1{{\smash{\mathop{\llra}\limits^{#1}}}}

\begin{document}

\title{Symmetries of the nucleon-nucleon $S$-matrix \\ and effective field theory expansions%\thanks{Grants or other notes
}


\author{Silas R.~Beane         \and
        Roland C.~Farrell %etc.
}


\institute{S.R.~Beane \and
           R.C.~Farrell \at
              Department of Physics,\\
              University of Washington, Seattle, WA 98195 \\
              \email{silas@uw.edu}, \email{rolanf2@uw.edu}
}

\date{Received: date / Accepted: date}

\maketitle

\begin{abstract}
  The s-wave nucleon-nucleon (NN) scattering matrix
  ($S$-matrix) exhibits UV/IR symmetries which are hidden in the
  effective field theory (EFT) action and scattering amplitudes, and
  which explain some generic features of the phase
  shifts. These symmetries offer clarifying interpretations of
  existing pionless EFT expansions, and suggest starting points for
  novel expansions. The leading-order (LO) $S$-matrix obtained in the
  pionless EFT with scattering lengths treated exactly is shown to
  have a UV/IR symmetry which leaves the sum of s-wave phase shifts
  invariant. A new scheme, which treats effective range corrections
  exactly, and which possesses a distinct UV/IR symmetry at LO, is
  developed up to NLO (next-to-LO) and compared with data.

\end{abstract}

\section{Introduction}
\label{intro}

\noindent The language of EFT was commonplace among particle physics
graduate students in the late 80's and early 90's. The syntax of this
language is quite simple and intuitive and amounts primarily to
counting dimensions of operators, and, for quantitative calculations,
including quantum effects perturbatively, as in the canonical example
of the chiral perturbation theory
expansion~\cite{Weinberg:1978kz,Gasser:1983yg}. This use of EFT
requires an understanding of regularization and renormalization at the
level of the textbook treatment of $\lambda\phi^4$ theory. Indeed, this
was covered in detail in Weinberg's two-year quantum field theory
sequence (attended by one of the authors) that took place several
times in that era and just preceded the initial publication of his
classic textbooks.

Contemporaneously, Weinberg's papers on EFT for nuclear physics
provided an organizational principle ---based in QCD and its
symmetries--- for establishing the most important terms in the nuclear
potential~\cite{Weinberg:1990rz,Weinberg:1991um,Weinberg:1992yk}.
However there was not much mention in these papers of the manner in
which the highly singular potentials would be regulated and
renormalized, or how such a procedure would
modify the organizational principle. The initial papers applying
Weinberg's ideas focused primarily on a direct numerical evaluation of
phase shifts to several non-trivial orders in the expansion and
largely ignored issues of renormalization by smoothing out singular
contributions to the potential using effective form
factors~\cite{Ordonez:1992xp,Ordonez:1993tn,Ordonez:1995rz}, the
standard practice in the nuclear physics community at the
time. Eventually, the issue of renormalization in Weinberg's scheme
was carefully considered in simple cases and rather dramatic
inconsistencies were found in the potentials describing the simplest
NN scattering phase shifts, initially in the s-wave
channels~\cite{Kaplan:1996xu}, and later in the p-wave
channels~\cite{Nogga:2005hy}. These observation spawned a research
program into the internal consistency of Weinberg's program which has
lasted over two decades and which still has achieved no
consensus\footnote{The literature is too vast to cite in its
  entirety. For helpful discussions, see
  Refs.~\cite{Beane:2001bc,PavonValderrama:2016lqn,vanKolck:2020llt}.}. In
parallel with this rather abstract theoretical research program, the
nuclear physics community has increasingly adopted the initial
Weinberg paradigm, while sometimes providing apologias for the
original renormalization issues; see,
e.g.~\cite{Epelbaum:2008ga,Machleidt:2011zz}.  Indeed, one could argue
that, while Weinberg's program has been pushed to ever increasing
sophistication and to larger and more complex nuclear systems, the
initial formal issues with consistent renormalization ---for the most fundamental systems (i.e. NN and NNN) which are
foundational to the entire program--- remain unresolved\footnote{For an alternate point of view that consistent renormalization has been achieved see Refs.~\cite{vanKolck:2020llt} and~\cite{Epelbaum:2017byx,Epelbaum:2017tzp,Epelbaum:2018zli,Epelbaum:2019msl,Epelbaum:2020maf,PavonValderrama:2019uzi}.}.

All of the above is in reference to the EFT relevant for
characteristic momenta $k\sim M_\pi$ and which therefore includes the
pion as a propagating degree of freedom in the EFT Lagrangian. Pions
introduce the complexities of a highly-singular tensor force, as well as
non-trivial aspects of the chiral symmetry of QCD. Efforts to resolve
the renormalization issues with Weinberg's program led to a new EFT
proposal in which pions are treated in perturbation theory and
consistent renormalization is
manifest~\cite{Kaplan:1998tg,Kaplan:1998we}.  While this scheme
ultimately failed to agree with the real world~\cite{Fleming:1999ee},
it was an important success as regards the change of mindset in
nuclear physics that was brought by Weinberg's work: it clearly
demonstrated that a specific EFT organizational principle, such as the one
proposed in Ref.~\cite{Kaplan:1998tg,Kaplan:1998we}, is falsifiable.

By contrast, inspired by Weinberg's work, tremendous progress was made
in understanding renormalization for $k\sim 1/a_s \ll M_\pi$, where
$a_s$ ($s=0,1$) are the s-wave NN scattering
lengths~\cite{vanKolck:1998bw,Kaplan:1998tg,Kaplan:1998we,Birse:1998dk}. At these
distance scales the pion and all its complexities can happily be
integrated out, leaving a relatively simple EFT with nucleon degrees
of freedom, generally interacting with some external
currents\footnote{Halo nuclei have also been described successfully by
  an extension of the pionless EFT
  technology~\cite{Bertulani:2002sz,Hammer:2017tjm}.}.  In the initial
studies of EFTs of NN scattering in which the potential is treated
exactly, a fundamental point of confusion lay in the belief that NN
scattering could be described by a coordinate-space potential given by
a gradient expansion of delta-function interactions. That this is not
the case was shown explicitly in Ref.~\cite{Beane:1997pk}, based on
earlier work which had clarified Wigner's causality bounds on
scattering parameters in the language of
EFT~\cite{Phillips:1996ae,Scaldeferri:1996nx,Phillips:1997xu,Hammer:2010fw}.  Long
ago Wigner showed that causality places constraints on scattering
parameters in quantum mechanical scattering with potentials of finite
range~\cite{Wigner:1955zz}.  In the EFT paradigm, in mass-dependent
schemes like cutoff regularization, the UV momentum-space cutoff acts
like the inverse of the range of interaction. While this relation is
somewhat fuzzy, removal of the cutoff in the renormalization procedure
is operationally equivalent to taking the range of interaction to
zero.  

It is worth mentioning here the flawed analogy between a
non-perturbative cutoff EFT of nuclear physics and EFTs of QCD
relevant for lattice QCD simulations.  While there is a clear parallel
between the lattice QCD action with Symanzik improvement and the
nuclear EFT Lagrangian, lattice QCD has a continuum limit which is
consistent with all general physical principles, including causality
while causality ---via the Wigner bound--- prohibits nuclear EFT from
achieving a continuum limit. This is no surprise physically: both
nuclear EFT and lattice QCD formally find their UV completion in QCD.
However, nucleonic degrees of freedom lose their meaning in the
approach to the continuum limit, and this is, at least qualitatively,
reflected in the Wigner bound in NN scattering\footnote{Note however
  that scattering with s-wave resonances can be described with
  zero-range forces~\cite{Habashi:2020qgw}.}. In some sense, the
Wigner bound pigeonholes nuclear EFT into the Wilsonian paradigm where
a cutoff on short-distance modes is implemented and then varied over
some range in order to demonstrate cutoff independence order-by-order
in the EFT.

The Wigner bound applies only when a strictly finite-range potential
is treated exactly, giving rise to a unitary $S$-matrix. Treating the
NN s-wave effective-range parameters to be of order the range of
interaction of the potential, allows the gradients of delta-function
potentials to be treated in perturbation
theory~\cite{vanKolck:1998bw,Kaplan:1998tg,Kaplan:1998we,Birse:1998dk}, and this
renders the Wigner bounds a non-issue. A highly efficient pionless
EFT, in which non-perturbative nuclear binding arises from treating
the momentum-independent four-nucleon operators exactly, while momentum-dependent operators are treated in perturbation theory, has
proven to be a remarkably successful paradigm for low-energy few-body
interactions with large scattering
lengths~\cite{vanKolck:1998bw,Kaplan:1998tg,Kaplan:1998we,Chen:1999tn,Hammer:2019poc}.
Of course, the inherent ambiguity in the form of the nuclear EFT
potential at short distance allows avoidance of the Wigner bound even
when the potential ---including momentum-dependent operators--- is
treated exactly~\cite{Gegelia:1998gn}, as long as the range of the
interaction is kept finite. However, as will be seen below, in this
case it is challenging to express amplitudes in a useful form, as
renormalization conditions have to be imposed at all orders in the
momentum expansion.

As the s-wave NN effective ranges are somewhat large as compared to
the pion's Compton wavelength, and indeed the pionless EFT expansion is
rendered more effective by expanding about the deuteron or dibaryon pole which
subsumes effective range corrections~\cite{Phillips:1999hh,Gegelia:2001ev,SanchezSanchez:2017tws}, it makes
sense to treat the range-corrections exactly~\cite{Kaplan:1999qa}
while continuing to treat shape-parameter effects in perturbation
theory.  A practical way to achieve this is by making use of the
dibaryon scheme developed for NN scattering in
Ref.~\cite{Kaplan:1996nv} and formulated in detail, with many
examples, in Ref.~\cite{Beane:2000fi}. This method effectively includes range
corrections non-perturbatively through the use of energy-dependent
potentials which, from the EFT perspective, can be constructed using operator
ambiguities and the freedom of field redefinition. One of the goals of
this contribution is to provide an EFT scheme which treats effective-range
corrections exactly using energy-independent interactions. It may
come as a surprise that there is a powerful symmetry which
constrains this system, and facilitates the avoidance of the Wigner
bound in a useful manner.

In the Wilsonian paradigm, EFTs are viewed as perturbations about
fixed points of the renormalization group (RG)~\cite{Birse:1998dk}. In
non-relativistic s-wave scattering with finite-range forces, the
running couplings exhibit two interesting fixed points, the trivial
fixed point where there is no interaction, and the unitary fixed point
where the interaction is as strong as it can be without violating
unitarity. At both fixed points, the EFT exhibits conformal invariance
or Schr\"odinger symmetry~\cite{Mehen:1999nd}.  Note however that the
notion that EFTs of contact forces are expansions about fixed points
of the RG is not strictly-speaking correct, as, at the fixed point
itself, there is no scattering. That is, the LO amplitude, with finite
scattering lengths, has an intrinsic scale and is therefore in itself
necessarily a perturbation about the RG fixed point.

The main point that will be conveyed in this contribution is that the
simplest $S$-matrices which describe s-wave scattering with
finite-range forces near (but not at) unitarity exhibit exact UV/IR
symmetries~\cite{Beane:2020wjl,Beane:2021bpr,Beane:2021xrk}.  A new
viewpoint here that may be unfamiliar to the reader is the focus on
the symmetries of the $S$-matrix, rather than the symmetries of the
action or the scattering amplitude.  In addition to providing a
new way of viewing the EFT near the unitary fixed point, UV/IR
symmetries suggest novel EFT expansions. In particular, the s-wave
scattering system at low energies, with scattering length and
effective range corrections treated exactly, is considered in detail
as an example of the power and utility of the UV/IR symmetries.

%%%%%%%%%%%%%%%%%%%%%%%%%%%%%%%%%%%%%%%%%%%%%%%%%%%%%%%%%%%%%%%%%%%%%%%%%%%%%%%%%%%%%%%%
\section{The $S$-matrix and UV/IR symmetry}
\label{sec:1}

\noindent Below inelastic thresholds, the unitary s-wave scattering amplitude for a finite range potential is
given by the effective range expansion (ERE)
\begin{equation}
T(k)\ =\ -\frac{4 \pi}{M}
\left[ -\frac{1}{a} +
\frac{1}{2} r {k^2} + v_{2} k^4 +v_{3} k^6 + O(k^8) - i k \right]^{-1},
\label{eq:reamp}
\end{equation}
where $k=\sqrt{ME}$ is the on-shell center-of-mass (c.o.m.) momentum, $a$ is the
scattering length, $r$ is the effective range, and the $v_{i}$ are shape parameters.
With the normalization of the scattering amplitude implied by Eq.~(\ref{eq:reamp}), the $S$-matrix element is given by
\begin{equation}
  S\ =\ e^{2 i \delta(k)}\ =\ 1 \ - \ i \frac{k M}{2\pi} T(k) \ .
   \label{eq:reampbS}
\end{equation}
Consider the scattering length approximation, where all effective range and shape parameters
are taken to vanish. In this case, the scattering amplitude and $S$-matrix element are
\begin{equation}
T(k)\ =\ \frac{4 \pi}{M} \frac{a}{ 1+i a k} \ \ \ , \ \ \ S(k) \ =\ \frac{1- i a k}{1+ i a k} \ .
\label{eq:reampscla}
\end{equation}
It is immediately clear that $S$ has a symmetry which is not a symmetry of $T$. The scale
transformation $k\mapsto e^\beta k$, $a\mapsto e^{-\beta} a$ leaves $S$ invariant and simply
implies that varying $k$ with $a$ fixed is the same as varying $a$ with $k$ fixed. The trivial
fixed point, $S=1$, is reached with no interaction, $a=0$ or with $k=0$, and the unitary fixed point,
$S=-1$, is reached with $a$ infinite or with $k$ infinite. 

Now consider the momentum inversion
\begin{eqnarray}
k\mapsto {1 \over a^2 k} \ .
   \label{eq:singlechanmominv}
\end{eqnarray}
The transformation on $T$ is complicated, however, it implies a simple transformation
of the $S$-matrix element
\begin{eqnarray}
  \delta(k) \mapsto -\delta(k) \pm \frac{\pi}{2} \ \ &,& \ \ S \to -S^* \ ,
\end{eqnarray}
where the sign of the $\pi/2$ phase is determined by the sign of the
scattering length.  This momentum inversion clearly interchanges the trivial
and unitary RG fixed points. As the trivial RG fixed point is, by definition, not seen by the
scattering amplitude, it is no surprise that the momentum inversion symmetry
does not act simply on $T$. The transformation interrelates the scattering amplitude and
the unit operator, and thus is realized on the $S$-matrix.

It may seem that the momentum inversion transformation of
Eq.~(\ref{eq:singlechanmominv}) is orthogonal to the idea of EFT since
it interchanges the UV and the IR. Note however that at LO in the
EFT, one is considering scattering in a limit in which all
short-distance mass scales are taken to be very large. In this
limit, long-distance forces (like pion exchange) and inelastic
thresholds (like the pion-production threshold) are only probed as
momentum approaches infinity. Therefore it is reasonable, at LO, to consider
transformations of the momenta over the entire momentum half-line, $0<k<\infty$. 

The fixed point of the momentum-inversion transformation\footnote{This
  fixed point is not to be confused with the fixed points of the RG at
  which the EFT exhibits Schr\"odinger symmetry and there is no
  scattering.} is the momentum that maps into itself under inversion,
$k^\circ=|a|^{-1}$, which is the absolute value of the scattering
amplitude's pole position in the complex momentum plane.  Therefore the EFT which reproduces the scattering length approximation is, strictly speaking, an
expansion about the fixed point, $k^\circ$, of the UV/IR
transformation.

Next, consider the inclusion of effective range effects with shape parameters
taken to vanish. In this case, the momentum inversion
\begin{eqnarray}
k\mapsto {2 \over |a r| k} \ ,
   \label{eq:singlechanmominv2}
\end{eqnarray}
implies the transformations of the $S$-matrix element
\begin{eqnarray}
  \delta(k) \mapsto -\delta(k) \ \ &,& \ \ S \to S^* \  \ , \ \ a r > 0 \ , \nn \\
  \delta(k) \mapsto \delta(k) \ \ &,& \ \ S \to S \  \ , \ \ a r < 0 \ .
\end{eqnarray}
The absence of a phase in this transformation reflects that, as a
function of momentum, the $S$-matrix element begins and ends at the
trivial RG fixed point, $S=1$.  This momentum inversion has the fixed
point $k^\circ=\sqrt{2}|a r|^{-1/2}$, which is the square root of the
absolute value of the product of the scattering amplitude's two pole
positions in the complex momentum plane. Again, if this amplitude with
vanishing shape parameters is treated as LO in an EFT, then the expansion
is about the scale set by
this fixed point.

%%%%%%%%%%%%%%%%%%%%%%%%%%%%%%%%%%%%%%%%%%%%%%%%%%%%%%%%%%%%%%%%%%%%%%%%%%%%%%%%%%%%%%%%
\section{UV/IR symmetries of the NN s-wave $S$-matrix}
\label{sec:uvir}

\noindent Realistic NN scattering at low energies is not a single-channel system as the initial-state nucleons can be arranged into two distinct spin configurations. The NN $S$-matrix at very low energies is dominated by the s-wave and can be written as~\cite{Beane:2018oxh,Beane:2020wjl}
\begin{eqnarray}
  \hat {\bf S}(\delta_0,\delta_1)
  & = &
{1\over 2}\left(  S_1+S_0 \right)
 \hat   {\bf 1}
\ +\
{1\over 2}\left( S_1-S_0 \right)
{\cal P}_{12} \ ,
  \label{eq:Sdef}
\end{eqnarray}
where now
\begin{equation}
  S_s\ =\ e^{2 i \delta_s(k)} \ ,
   \label{eq:reampbSsingle}
\end{equation}
the SWAP operator is
\begin{eqnarray}
{\cal P}_{12} = \oneht \left({ \hat   {\bf 1}}+  \hat  {\bm \sigma} \cdot   \hat  {\bm \sigma} \right)
  \label{eq:swap}
\end{eqnarray}
and, in the direct-product space of the nucleon spins,
\begin{eqnarray}
 \hat {\bf 1} \equiv \hat {\cal I}_2\otimes  \hat {\cal I}_2 \ \ \ , \ \
  \hat {\bm \sigma} \cdot \hat {\bm \sigma} \equiv \sum\limits_{\alpha=1}^3 \
\hat{ \sigma}^\alpha \otimes \hat{ \sigma}^\alpha \ ,
  \label{eq:Hil}
\end{eqnarray}
with ${\cal I}_2$ the $2\times 2$ unit matrix, and $\hat{
  \sigma}^\alpha$ the Pauli matrices. The $\delta_s$ are s-wave phase
shifts with $s=0$ corresponding to the spin-singlet ($\si$) channel
and $s=1$ corresponding to the spin-triplet ($\siii$) channel. Below,
the following physical effective range parameters will be used~\cite{Kaplan:1998we,deSwart:1995ui}: $a_{0}
= -23.714(13)$ fm, $a_{1} = 5.425(1)$ fm, $r_{0} = 2.73(3)$ fm, and
$r_{1} = 1.749(8)$ fm. 

The momentum-inversion transformations of the $S$-matrix
elements considered above are not transformations that act simply on
the full $S$-matrix as they are s-wave channel dependent. However,
these transformations can be readily generalized to the full
$S$-matrix. In the scattering length approximation, consider the
momentum inversion transformation~\cite{Beane:2020wjl}
\begin{eqnarray}
p\mapsto \frac{1}{|a_1 a_0| p} \ .
   \label{eq:slafullsmat}
\end{eqnarray}
This implies the following transformations on the phase shifts for the NN physical case $a_1 a_0 <0$:
\begin{eqnarray} 
&&   \delta_0(p) \mapsto \delta_1(p) - \frac{\pi}{2} \ \ \ , \ \ \ \ \ \, \delta_1(p) \mapsto \delta_0(p) + \frac{\pi}{2} \ .
  \label{eq:confmoebiusiso2}
\end{eqnarray}
This transformation therefore corresponds to an exact symmetry, which leaves the combination of phase shifts $\delta_0+\delta_1$ invariant.
As phase shifts are angles, this is, by definition, a conformal invariance, albeit one that allows the presence of a mass scale. This
conformal invariance facilitates the construction of a purely geometrical theory of scattering~\cite{Beane:2020wjl,Beane:2021bpr}.

With effective range corrections included, the momentum inversion transformation
\begin{eqnarray}
p\mapsto \frac{1}{\lambda |a_1 a_0| p} \ ,
   \label{eq:moebiusgen}
\end{eqnarray}
with the arbitrary real parameter $\lambda >0$, implies
\begin{eqnarray} 
&&   \delta_0(p) \mapsto \delta_0(p) \ \ \ , \ \ \ \ \ \, \delta_1(p) \mapsto -\delta_1(p) \ ,
  \label{eq:confmoebiusiso3}
\end{eqnarray}
but only in the special case\footnote{The general case is considered in Ref.~\cite{Beane:2021bpr}.} where the effective ranges are correlated with the scattering lengths 
as 
\begin{eqnarray} 
r_0&=&2\lambda a_1 \ \ \ ,\ \ \ r_1\,=\,-2\lambda a_0\ .
  \label{eq:confmoebiusiso5}
\end{eqnarray}

This UV/IR symmetry has interesting implications for
nuclear physics. The measured singlet NN scattering phase shift rises
steeply from zero due to the unnaturally large scattering length, and
then, as momenta approach inelastic threshold, the phase shift goes
through zero and becomes negative, indicating the fabled short-distance
repulsive core. While this impressionistic description assigns
physics to the potential, which is not an observable and indeed need
not be repulsive at short distances, the UV/IR symmetry directly
imposes the physically observed behavior of the phase shift. Even though the
singlet phase shift changes sign at momenta well beyond the range of
applicability of the pionless theory, ascribing this symmetry to the
LO results ---through the required presence of range corrections--- results
in a more accurate LO prediction than the usual pionless expansion~\cite{SanchezSanchez:2017tws}.

%%%%%%%%%%%%%%%%%%%%%%%%%%%%%%%%%%%%%%%%%%%%%%%%%%%%%%%%%%%%%%%%%%%%%%%%%%%%%%%%%%%%%%%%
\section{EFT description: single-channel case}

\subsection{Potential and Lippmann-Schwinger equation}

\noindent It is convenient to introduce the generic UV scale ${\cal M}$ and the
generic IR scale $\aleph$. The EFT will describe physics for $k\sim \aleph \ll {\cal M}$.
Subleading corrections to the scattering amplitude in the EFT are expected to be parametrically
suppressed by powers of $k/{\cal M}$. For the NN system at very-low energies, described by the pionless EFT, the UV scale is ${\cal M}\sim M_\pi$.

The s-wave potential stripped from the most general effective Lagrangian of four-nucleon contact operators is~\cite{Kaplan:1996nv,Phillips:1997xu}
\begin{equation}
V(p',p)=C_0 + C_2 \left(p^2 + p'^2\right) + C_4 \left(p^4 + p'^4\right) + C_4' p^2 p'^2+\dots \ ,
\label{eq:Vexp}
\end{equation}
where the $C_n$ are the bare coefficients.
The scattering amplitude is obtained by solving the Lippmann-Schwinger (LS) equation with this potential
\begin{equation}
T(p',p;E)=V(p',p) + M\int \frac{d^3q}{(2 \pi)^3} \, V(p',q) 
\frac{1}{EM- {q^2}+i\epsilon} T(q,p;E) \ .
\label{eq:LSE}
\end{equation}
As the potential is separable to any order in the momentum expansion, the scattering amplitude can be
obtained in closed form to any desired order in the
potential~\cite{Phillips:1997xu}. Of course the singular nature of the
potential requires regularization and renormalization. The scattering
amplitude can be regulated using, for instance, dimensional regularization and its various schemes,  or by
simply imposing a hard UV cutoff, $\Lambda$, on the momentum integrals.

\subsection{Matching equations}

\noindent In the language of cutoff regularization, matching the solution of the LS equation with the ERE of
Eq.~(\ref{eq:reamp}), formally gives the all-orders matching equations
\begin{eqnarray}
a & =& f_0(\Lambda; C_0,C_2,C_4,\ldots) \nn \ , \\
r  & =& f_2(\Lambda; C_0,C_2,C_4,\ldots) \nn \ , \\
v_n  & =& f_{2n}(\Lambda; C_0,C_2,C_4,\ldots) \ ,
  \label{eq:match}
\end{eqnarray}
where the $f_{2n}$ are non-linear functions determined by solving the
LS equation.  These equations can be inverted to find the now
cutoff-dependent coefficients $C_{2n}(\Lambda;a,r,v_2,\ldots)$.  To
obtain an EFT with predictive power, it is necessary to identify the
relative size of the effective range parameters. If they all scale as
powers of ${\cal M}^{-1}$, then the entire potential of
Eq.~(\ref{eq:Vexp}) can be treated in perturbation theory for momenta
$k\ll {\cal M}$. If there are big parts that instead scale as powers
of ${\aleph}^{-1}$, they will, via the interactions of Eq.~(\ref{eq:Vexp}), constitute the LO potential in the EFT
expansion.

\subsection{Scattering length approximation}

\noindent With $a\sim \aleph^{-1}$ and the effective range and shape parameters of 
natural size, $r\sim {\cal M}^{-1}$, $v_n\sim {\cal M}^{-2n+1}$, 
the amplitude can be expanded for $k\ll {\cal M}$ as
\begin{equation}
T(k)\ =\ -\frac{4 \pi}{M}\left( -\frac{1}{a} - i k \right)^{-1} \bigg\lbrack 1 + O(k) \bigg\rbrack \ .
\label{eq:reampb3}
\end{equation}
In order to generate this expansion in the EFT, the potential is written as
\begin{equation}
V(p',p)=V_{\LOss}+V_r(p',p) \ ,
\label{eq:Vexp2}
\end{equation}
where $V_{\LOss}=C_0$ is treated exactly in the LS equation and the residual potential, $V_r$, which includes range
and shape parameter corrections, is treated in perturbation theory.
Keeping the first term in the s-wave potential, the solution of the LS equation is
\begin{equation}
T_{\LOss}(k) =\left( \frac{1}{C_0} - {\mathbb{I}}(k) \right)^{-1},
\label{eq:sla1}
\end{equation}
where 
\begin{equation}
\mathbb{I}(k) \equiv  \left(\frac{\omega}{2}\right)^{3-d}  M\int \frac{d^dq}{(2 \pi)^{d}} \, \frac{1}
{{k^2}- {q^2}+i\epsilon}\ \  \mapup{\rm PDS}\ \ -\frac{M}{4\pi}\left(\omega +i k \right) \ ,
\label{eq:IEdef}
\end{equation}
has been evaluated in dimensional regularization with the PDS scheme~\cite{Kaplan:1998tg,Kaplan:1998we} and renormalized at the RG scale $\omega$.
The matching equations in this case are
\begin{eqnarray}
&& a = \left(\frac{4\pi}{M C_0} + \omega \right)^{-1}  \ ,\nn \\
&& r = v_n  = 0 \ .
  \label{eq:matchsla}
\end{eqnarray}
Inverting one finds
\begin{equation}
  C_0(\omega) \ =\ \frac{4 \pi}{M} \frac{1}{{1}/{a}-{\omega}} \ .
\end{equation}
The coupling at the unitary fixed point, $C_0=C_{0\star}$, corresponds to a divergent scattering
length (unitarity). A rescaled coupling can be defined as
${\hat C}_0 \equiv C_0/C_{0\star}$. The corresponding beta-function 
is then 
\begin{equation}
  {\hat\beta({\hat C}_0)}\ =\  \omega \frac{d}{d\omega} {\hat C}_0(\omega) \ =\ -{\hat C}_0(\omega)\left({\hat C}_0(\omega)-1\right) \ ,
   \label{eq:c0betafn}
\end{equation}
which explicitly has fixed points at ${\hat C}_0=0$ and $1$. Now note that the beta-function is invariant
with respect to the inversion of the RG scale~\cite{Beane:2021xrk}
\begin{eqnarray}
\omega \mapsto {1 \over a^2 \omega} \ ,
   \label{eq:singlechancutoffinv}
\end{eqnarray}
which interchanges the trivial and unitary fixed points
\begin{equation}
  {\hat C}_0(\omega) \leftrightarrow 1 - {\hat C}_0(\omega) \ .
     \label{eq:chattransf}
\end{equation}
This exactly mirrors the action of the UV/IR transformation on the $S$-matrix elements as was shown in section~\ref{sec:1},
and shares the fixed point $\omega^\circ=k^\circ=|a|^{-1}$.
The moral here is that while the UV/IR transformation is not simply realized at the level of the EFT action, its effect is manifest in the RG flow of the EFT operators~\cite{Beane:2021xrk}.

\subsection{Range corrections with zero-range forces}

\noindent With $a, r\sim \aleph^{-1}$ and shape parameters of 
natural size, $v_n\sim {\cal M}^{-2n+1}$, 
the amplitude can be expanded for $k\ll {\cal M}$ as
\begin{equation}
T(k)\ =\ -\frac{4 \pi}{M}\left( -\frac{1}{a} + \frac{1}{2} r {k^2} - i k \right)^{-1} \bigg\lbrack 1 + O(k^3) \bigg\rbrack \ .
\label{eq:reampb2}
\end{equation}
In order to generate this expansion from the EFT perspective, one may choose $V_{\LOss}=C_0+C_2
\left(p^2 + p'^2\right)$. Due to the highly singular UV behavior of this potential, cutoff
regularization will be used to carefully account for the divergences that are generated when
solving the LS equation. The amplitude is found in closed form to be~\cite{Phillips:1997xu}
\begin{equation}
T_{\LOss}(k)=\left(\frac{(C_2 I_3 -1)^2}{C + C_2^2 I_5 + {k^2} C_2 (2 - C_2 I_3)} - {\mathbb I}(k)\right)^{-1} \ ,
\label{eq:Tonexp}
\end{equation}
where now ${\mathbb I}(k)$ is evaluated with cutoff regularization, and
\begin{equation}
I_n \equiv -M \int \frac{d^3q}{(2 \pi)^3} q^{n-3}\theta \left(\frac{\pi}{2}\Lambda-q\right)=-\frac{M \Lambda^n}{2^{1+n}\pi^{2-n}n}\ .
\label{eq:In}
\end{equation}
Matching the amplitude to the ERE gives
\begin{eqnarray}
a &=& \frac{M}{4 \pi}\left(\frac{(C_2 I_3 -1)^2}{C + C_2^2 I_5} - {I_1}\right)^{-1} \ ,\nn \\
r &=& \frac{8 \pi}{M}\left(\frac{(C_2 I_3 -1)^2}{C + C_2^2 I_5}\right)^2
\left[\frac{1}{(C_2 I_3 - 1)^2 I_3} - \frac{1}{I_3}\right]    \ ,   \nn \\
v_n  &=&  -\frac{4 \pi}{M}\frac{C_2^n(C_2 I_3 -2)^n(C_2 I_3 -1)^2}{(C + C_2^2 I_5)^{n+1}} \ .
  \label{eq:matchcrm}
\end{eqnarray}
Taking the limit $\Lambda \rightarrow \infty$ then recovers the ERE
with all shape parameter corrections vanishing, $v_n=0$. However, it
is straightforward to verify that in this limit, $r \leq 0$, as
required by the Wigner bound~\cite{Wigner:1955zz}. If s-wave NN scattering involved
negative effective ranges and resonances rather than bound states,
then this scheme, with the effective potential consisting of a finite
number of strictly delta-function potentials, would
suffice~\cite{Habashi:2020qgw}. However, as the s-wave NN effective ranges are
both positive, it is clear that $\Lambda$ must be kept
finite. In that case the higher-order terms in the bare
potential, even if neglected in the choice of $V_{\LOss}$, are generated
quantum mechanically at LO as evidenced by the non-vanishing shape
parameters. Therefore the higher order operators should be kept from the start. That is, one
is back to the general, formal statement of the matching conditions
given in Eq.~(\ref{eq:match}), where a renormalization scheme is
required which ensures that $v_n=0$, and a choice of $V_{\LOss}$ must be
found which achieves this while including all orders in the momentum
expansion.

\subsection{Range corrections with a finite-range scheme: LO}

\noindent The lesson provided by the Wigner bound is that if range corrections are treated exactly, then the LO potential
should, in general, include all orders in the momentum expansion. Therefore, the potential
can be written as in Eq.~(\ref{eq:Vexp2}) except now with a momentum dependent LO potential
\begin{equation}
V(p',p)=V_{\LOss}(p',p) + V_r(p',p) \ .
\label{eq:VexpEF}
\end{equation}
The residual potential, $V_r$, which accounts for NLO and higher effects in
perturbation theory, will be considered in detail below.
Now the potential is non-unique and there is no reason for the
separation into $V_{\LOss}$ and $V_r$ to be unique. Ideally one
finds a LO potential which identically gives the ERE with all
shape parameters vanishing, and indeed that is what will be achieved
by imposing the UV/IR symmetry at the level of the interaction.

As the s-wave EFT potential is non-local (non-diagonal in coordinate space) and separable to any order in the momentum expansion, it appears sensible to assume that the LO potential which generates the ERE with range corrections only is non-local and separable. However, such an assumption is not necessary: all potentials which generate the ERE truncated at the effective
range are, by definition, phase equivalent\footnote{Note that the potential
considered in the previous section is phase equivalent only in the limiting sense of
$\Lambda \to \infty$ and $r < 0$.}. Once one potential is found, others can be
obtained by unitary transformation. Here the simplest possibility
will be considered; that the LO potential is (rank-one) separable.

For a separable potential $V(p,p')$, the on-shell scattering amplitude
solves the LS equation algebraically to
\begin{equation}
T(k)\ =\ {V(k,k)}\left(1-M\int \frac{d^3q}{(2 \pi)^{3}} \, \frac{V(q,q)}
{{k^2}- {q^2}+i\epsilon}\right)^{-1} \ .
\label{sepsol}
\end{equation}
The momentum inversion symmetry of Eq.~(\ref{eq:singlechanmominv2})
(assuming for simplicity $\eta\equiv a r/2 >0$) implies
\begin{equation}
  T(k)\mapsto T\left({1 \over {\eta k}}\right) \ =\ -\eta k^2\, T^*(k) \ .
   \label{sepsol2}
\end{equation}
With the range of the potential taken to be
the momentum scale $\mu$, the potential can be taken to be the real function $V_{\LOss}(\mu;p,p')$. Constraints on the $S$-matrix concern the potential
$V_{\LOss}(\mu;p,p)\equiv V_{\LOss}(\mu;p)$. Therefore,
\begin{eqnarray}
  \hspace{-0.3in}T^*(k) & =&
{V_{\LOss}\left(\mu;k\right)}\left({1-M\int \frac{d^3q}{(2 \pi)^{3}} \, \frac{V_{\LOss}(\mu;q)} {{k^2}- {q^2}-i\epsilon}}\right)^{-1} \nn \\
     & =&
     \frac{-\frac{1}{\eta k^2} V_{\LOss}\left(\mu;{1 \over {\eta k}}\right)}
     {\bigg\lbrack1-\frac{M}{2\pi^2}\int dq \, \left(-\frac{1}{\eta q^2}\right){V_{\LOss}(\mu;{1\over {\eta q}})}\bigg\rbrack
-M\int \frac{d^3q}{(2 \pi)^{3}} \,  \left(-\frac{1}{\eta q^2}\right)\frac{V_{\LOss}(\mu;{1\over {\eta q}})} {{k^2}- {q^2}-i\epsilon}}
       \label{eq:LSgen}
\end{eqnarray}
where the first line follows from Eq.~(\ref{sepsol}) and the second line follows from Eq.~(\ref{sepsol2}).
Now consider non-singular solutions of this equation of the form
\begin{equation}
V_{\LOss}\left(\mu;{1 \over {\eta q}}\right)\ = \ \epsilon\, \eta q^2\; V_{\LOss}(\nu;q)\ ,
\label{eq:yamatransfgen}
\end{equation}
where $\epsilon=\pm 1$, and the range of the potential has been allowed to vary under momentum inversion,
and so $\nu$ is an independent scale in correspondence with the range of the transformed potential. Integrating this equation gives
\begin{equation}
\int_0^\infty dq\, V_{\LOss}\left(\mu;q\right)\ = \ \epsilon\,\int_0^\infty dq\, V_{\LOss}(\nu;q)\ .
\label{eq:intcongen}
\end{equation}
With the choice $\epsilon=-1$ and $\mu=\nu$, this integral must vanish identically and there is clearly a formal solution
to Eq.~(\ref{eq:LSgen}). However, an explicit solution does not seem to exist for a real, finite-range potential.

What follows focuses on the case $\epsilon=1$ and $\mu\neq \nu$. In this case, Eq.~(\ref{eq:yamatransfgen}) is solved by
the ratio of polynomials
\begin{equation}
  V^n_{\LOss}\left(\mu;p\right)\ = \ \frac{N}{M\mu}\frac{1-\left(\frac{p^2}{\mu^2}\right)^{n+1}}{1-\left(\frac{p^2}{\mu^2}\right)^{n+2}} \ ,
\label{eq:gensolinv}
\end{equation}
where $n\in \mathbb{Z}: n\geq 0$,  $N$ is a dimensionless normalization constant and $\mu\,\nu=1/\eta$. One finds
\begin{equation}
\int_0^\infty dq\, V^n_{\LOss}\left(\mu;q\right)\ = \ \int_0^\infty dq\, V^n_{\LOss}\left(\nu;q\right)\ = \ \frac{\pi  N}{M (n+2)} \cot \left(\frac{\pi }{2 n+4}\right) \ .
\label{eq:intcongen2}
\end{equation}
With the postulated solution, Eq.~(\ref{eq:LSgen}) takes the form
\begin{eqnarray}
  \hspace{-0.3in}T^*(k) & =&
{V^n_{\LOss}\left(\mu;k\right)}\left({1-M\int \frac{d^3q}{(2 \pi)^{3}} \, \frac{V^n_{\LOss}(\mu;q)} {{k^2}- {q^2}-i\epsilon}}\right)^{-1} \nn \\
& =&
{{\bar V}^n_{\LOss}\left(\nu;k\right)}\left({1-M\int \frac{d^3q}{(2 \pi)^{3}} \, \frac{{\bar V^n}_{\LOss}(\nu;q)} {{k^2}- {q^2}-i\epsilon}}\right)^{-1} \ ,
       \label{eq:LSgenspec}
\end{eqnarray}
where
\begin{eqnarray}
  {\bar V}^n_{\LOss}\left(\nu;k\right)&=& -{V}^n_{\LOss}\left(\nu;k\right)\left( 1 + \frac{M}{2\pi^2}\int_0^\infty dq\,V^n_{\LOss}\left(\nu;q\right)\right)^{-1} \nn\\
&=& -\frac{N}{M\nu\left(1+\frac{  N}{2\pi (n+2)} \cot \left(\frac{\pi }{2 n+4}\right)\right)}\left(\frac{1-\left(\frac{k^2}{\nu^2}\right)^{n+1}}{1-\left(\frac{k^2}{\nu^2}\right)^{n+2}}\right) \ .
       \label{eq:PEpot}
\end{eqnarray}
Now all that is required is to show that $V^n_{\LOss}(\mu;k)$ and ${\bar V}^n_{\LOss}(\nu;k)$ are phase-equivalent potentials. Equating
the two sides of Eq.~(\ref{eq:LSgenspec}) at both $k=0$ and $k = \sqrt{\mu \nu}$
yields a single solution for $n$ and $N$:
\begin{equation}
n=0 \ \ \ , \ \ \ N= -4\pi\left(1 + \frac{\nu}{\mu}\right)\ .
\label{erepars2}
\end{equation}
Solving the LS equation with either phase-equivalent potential gives
\begin{equation}
a= \frac{1}{\mu}+\frac{1}{\nu}  \ \ , \ \ {r}=\frac{2}{\mu+\nu} \ \ ,\ \ v_n=0 \ .
\label{erepars}
\end{equation}
Finally, the potential takes the separable and non-local form~\cite{Yamaguchi:1954mp,Beane:1997pk,Phillips:1999bf}
\begin{equation}
  V_{\LOss}\left(\mu,\nu;p,p'\right)\ = \ -\frac{4\pi}{M\mu}\left(1 + \frac{\nu}{\mu}\right)\frac{1}{  \sqrt{1+\frac{p^2}{\mu^2}}\sqrt{1+\frac{p^{\prime 2}}{\mu^2}}   } \ ,
\label{eq:gensolinvyam}
\end{equation}
with phase-equivalent potential $\bar V_{\LOss}\left(\mu,\nu;p,p'\right)=V_{\LOss}\left(\nu,\mu;p,p'\right)$. 

Note that the potential of Eq.~(\ref{eq:gensolinvyam}) satisfies the scaling law of Eq.~(\ref{eq:yamatransfgen}) only if $N$ is held fixed. In some sense, the original potential that appears in the
LS equation may be viewed as a bare potential which is determined by requiring that Eq.~(\ref{eq:yamatransfgen}) leave the LS form invariant. The corresponding transformation of the $S$-matrix, $S \to S^*$, is seen only after solving the LS
equation, which fixes $N$ to its $\mu$ and $\nu$ dependent value in Eq.~(\ref{erepars2}). $N$ is therefore a kind of anomalous scaling factor which takes the bare potential to the renormalized form of Eq.~(\ref{eq:gensolinvyam}). Letting the UV/IR transformation act on both the momenta, $p$, and the scales, $\mu$ and
$\nu$, (i.e. $\mu \leftrightarrow \nu$), generates a transformation on $V_{\LOss}$ ($\bar V_{\LOss}$) which is augmented from that of Eq.~(\ref{eq:yamatransfgen}) by an anomalous
scaling factor of $\nu/\mu$ ($\mu/\nu$). One important observation is that, up to an anomalous scaling factor and a sign, $V_{\LOss}$ and $\bar V_{\LOss}$ transform in the same manner as the amplitude that they generate, Eq.~(\ref{sepsol2}). 

With $\mu$ and $\nu$ intrinsically positive, the general case, with scattering length and effective range of any sign
is obtained by taking $\zeta\mu\,\nu=1/\eta$, with $\zeta=1$ corresponding to $a r >0$ and $\zeta=-1$ corresponding to $a r <0$.
The general solution is then
\begin{equation}
  V_{\LOss}\left(\mu,\nu;p,p'\right)\ = \ -\frac{4\pi}{M\mu}\left(1 + \zeta\frac{\nu}{\mu}\right)\frac{1}{  \sqrt{1+\frac{p^2}{\mu^2}}\sqrt{1+\frac{p^{\prime 2}}{\mu^2}}   } \ ,
\label{eq:gensolinvyamanysing}
\end{equation}
with phase-equivalent potential  $\bar V_{\LOss}\left(\mu,\nu;p,p'\right)=\zeta V_{\LOss}\left(\nu,\mu;p,p'\right)$, and
\begin{equation}
a= \frac{1}{\mu}+\frac{1}{\zeta \nu}  \ \ , \ \ {r}=\frac{2}{\mu+ \zeta \nu} \ \ ,\ \ v_n=0 \ .
\label{ereparsgen}
\end{equation}

Having both $a$ and $r$ large as compared to the (inverse) UV scale
${\mathcal M}^{-1}\sim M^{-1}_\pi$ generally requires $\mu,\nu\sim \aleph$.
Expanding
$V_{\LOss}$ in powers of the momenta for $p,p'\ll\aleph$ and matching
onto the momentum expansion of Eq.~(\ref{eq:Vexp}) leads to the scaling
\begin{eqnarray}
C^{(\prime)}_{\scriptscriptstyle{LO}\,m} \ \sim \ \frac{4\pi}{M \aleph^{m+1}} \ .
\label{eq:bcpmodel3}
\end{eqnarray}
The coefficients of the residual potential 
are expected to be suppressed, in a manner to be determined below, by the UV scale. As
the potential is not unique, the decomposition into IR enhanced and UV
suppressed contributions is not unique.  Treating the expanded LO
potential as a renormalization scheme, then for momenta $p,p'\sim
\aleph$, all $C^{(\prime)}_{\scriptscriptstyle{LO}\,m}$ terms in the
potential should be summed into the LO potential to give
Eq.~(\ref{eq:gensolinvyamanysing}) which is treated exactly in the LS
equation, while the residual potential is treated in perturbation theory.
\begin{figure*}
 \centerline{\includegraphics[width=0.7\textwidth]{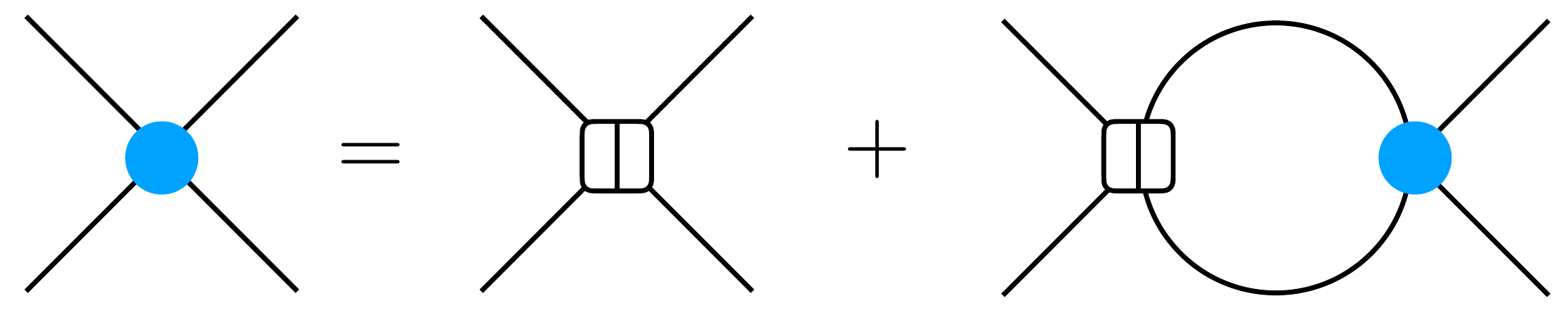}}
\caption{The LO scattering amplitude from the LS equation. The blue circle corresponds to the LO amplitude and the split square is an insertion of the LO separable potential.}
\label{fig:DiaLO}       % Give a unique label
\end{figure*}

\subsection{Range corrections with a finite-range scheme: NLO}

\noindent Recall that treating $a,r\sim \aleph^{-1}$ and $v_n\sim {\mathcal M}^{-2n+1}$, for $k\ll {\mathcal M}$, the ERE of Eq.~(\ref{eq:reamp}) can be expanded to
give the NLO amplitude 
\begin{equation}
T_{\NLOss}(k)\ =\ \frac{4 \pi}{M}\left( -\frac{1}{a} + \frac{1}{2} r {k^2} - i k \right)^{-2} v_2 k^4 \ .
\label{eq:eftNLOrr}
\end{equation}
Note that it has been assumed in this expression that $r$ is close to its ``physical'' value.
More generally, and of greater utility when considering realistic NN scattering, one can decompose $r=r_{\LOss}+r_{\NLOss}$, where $r_{\LOss}\sim \aleph^{-1}$ and $r_{\NLOss}\sim {\mathcal M}^{-1}$. In this case
\begin{equation}
T_{\NLOss}(k)\ =\ \frac{4 \pi}{M}\left( -\frac{1}{a} + \frac{1}{2} r_{\LOss} {k^2} - i k \right)^{-2} \oneht r_{\NLOss} k^2
\label{eq:eftNLOrr2}
\end{equation}
so that shape-parameter corrections enter at NNLO (as a subleading contribution to range-squared effects).

\begin{figure*}
 \centerline{\includegraphics[width=0.97\textwidth]{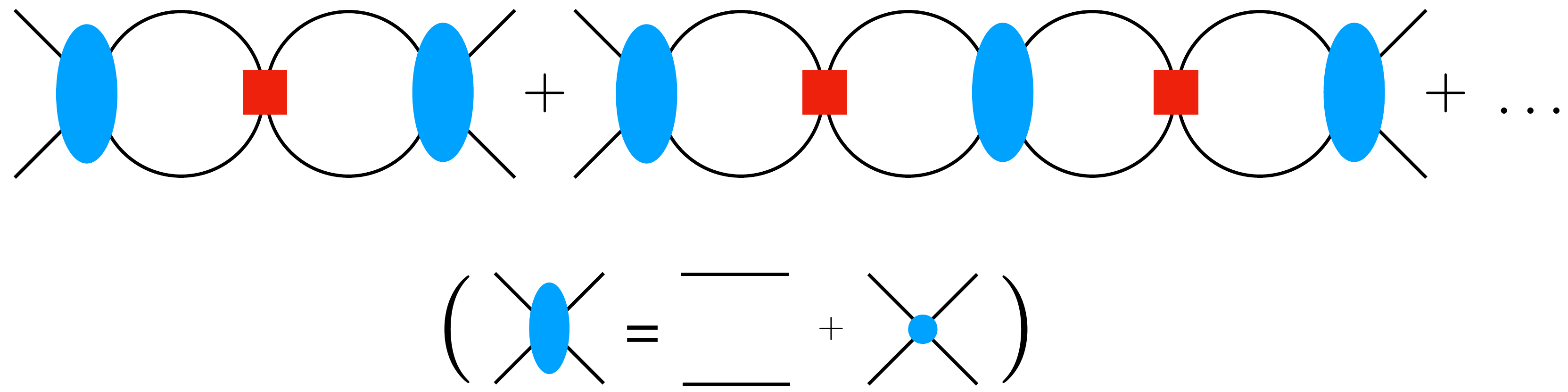}}
\caption{The NLO and beyond scattering amplitude in perturbation theory. The blue circle corresponds to the LO amplitude and the red square is an insertion of the NLO potential.}
\label{fig:DiaNLO}       % Give a unique label
\end{figure*}

The goal in what follows is to generate the NLO amplitudes of Eq.~(\ref{eq:eftNLOrr2}) and Eq.~(\ref{eq:eftNLOrr}) (in that order) in the EFT.
The LS equation, Eq.~(\ref{eq:LSE}), for the full scattering amplitude is symbolically expressed as
\begin{equation}
T\ = \ V +  V\, G\, T \ ,
\label{eq:LSE2}
\end{equation}
where $G$ is the two-particle Green's function. Expanding the scattering amplitude and potential as
\begin{equation}
T=T_{\LOss}+T_{\NLOss}+\ldots \ \ \ ,\ \ \  V=V_{\LOss}+V_r
\label{eq:TandVexp}
\end{equation}
leads to $T_{\LOss}$ as an exact solution of the LS equation, as illustrated in Fig.~\ref{fig:DiaLO}, and the NLO and beyond amplitude
\begin{equation}
T_{\NLOss}\ +\ \ldots =V_r + V_r\, G\, T_{\LOss} + T_{\LOss}\, G\, V_r + T_{\LOss}\, G\, V_r\, G\, T_{\LOss} \ \ldots \ ,
\label{eq:LSE3}
\end{equation}
as illustrated diagrammatically to NLO via the Feynman diagrams in Fig.~\ref{fig:DiaNLO}. The form of
the bare EFT potential $V_r$ which matches to the expanded ERE is
straightforward to find using the UV/IR symmetry.  It is convenient to express the LO potential in the compact form
\begin{equation}
  V_{\LOss}\left(p,p'\right)\ = \ C_{\scriptscriptstyle{LO}\,0}\, {\mathcal G}(p) {\mathcal G}(p') \ .
\label{eq:LOcompact}
\end{equation}
where $C_{\scriptscriptstyle{LO}\,0}$ and ${\mathcal G}(p^{(\prime)})$ are defined by comparing with Eq.~(\ref{eq:gensolinvyamanysing}).
The LO amplitude is then
\begin{equation}
  T_{\LOss}\left(k\right)\ = \ C_{\scriptscriptstyle{LO}\,0}\, {\mathcal G}^2(k)\,{\bm Z}^{-1} \ ,
\label{eq:LOcompact2}
\end{equation}
with ${\bm Z} \equiv 1-C_{\scriptscriptstyle{LO}\,0}{\mathbb I}_2$ and the convergent integral 
\begin{equation}
{\mathbb I}_2 \ =\ M\int \frac{d^3q}{(2 \pi)^3} \frac{{\mathcal G}^2(q)}{k^2- {q^2}+i\epsilon} \ =\ -\frac{M}{4\pi}{\mathcal G}^2(k)\left(\mu+i k\right) .
\label{eq:LOcompact3}
\end{equation}

Now notice that the NLO amplitude of Eq.~(\ref{eq:eftNLOrr2}) transforms simply under the momentum inversion $k\mapsto 2/( \vert a r \vert k)= \mu\nu/k$ as $T_{\NLOss}\to
T^*_{\NLOss}$ $(a r > 0)$ or $T_{\NLOss}\to T_{\NLOss}$ $(a r <
0)$. Based on the discussion below Eq.~(\ref{eq:gensolinvyam}) it may be expected that the part of $V_r$ that generates
$T_{\NLOss}$ will be invariant under momentum inversion up to a sign and $\mu \leftrightarrow \nu$.
Consider the energy-dependent potential,
\begin{equation}
k^2  V_{\LOss}\left(k,k\right){\mathcal G}^2(k)\ = \ C_{\scriptscriptstyle{LO}\,0}k^2\, {\mathcal G}^4(k)
\label{eq:LOcompact4}
\end{equation}
which maps to (minus) itself with $\mu\leftrightarrow\nu$ for $a r$ positive (negative) and is therefore a candidate for the piece of $V_r$ which generates $T_{\NLOss}$. An energy-independent residual potential can then be defined as
\begin{equation}
V_r(p',p)= \big\lbrack c_0+ c_2 \left(p^2 + p'^2\right)+\dots \big\rbrack V_{\LOss}\left(p',p\right)\left({\mathcal G}^2(p)+{\mathcal G}^2(p')\right)\ ,
\label{eq:VexpEF2}
\end{equation}
where the $c^{(\prime)}_m$ coefficients are bare parameters, subject to renormalization. On-shell this potential has the desired UV/IR transformation properties. Note that the form of the potential reflects that only odd powers of ${\mathcal G}(p^{(\prime)})$ in $V_r$ will match to
a polynomial in $k$ for $k\sim \aleph$. Formally, 
the (bare) coefficients of the full potential can be expressed for $k\ll \aleph$ as 
$C^{(\prime)}_m=C^{(\prime)}_{\scriptscriptstyle{LO}\,m}+C^{(\prime)}_{\scriptscriptstyle{NLO}\,m}$, where $C^{(\prime)}_{\scriptscriptstyle{NLO}\,m}\equiv C_{\scriptscriptstyle{LO}\,0}\cdot c^{(\prime)}_m$.

Evaluating the diagrams of Fig.~\ref{fig:DiaNLO} with a single insertion of $V_r$ gives
\begin{eqnarray}
&& T_{\NLOss} = 2c_0 C_{\scriptscriptstyle{LO}\, 0} {\mathcal G}^2(k)\, \big\lbrack {\mathcal G}^2(k) + C_{\scriptscriptstyle{LO}\, 0}{\bm Z}^{-1}\left({\mathcal G}^2(k){\mathbb I}_2+{\mathbb I}_4\right) 
+ C^2_{\scriptscriptstyle{LO}\, 0}{\bm Z}^{-2}{\mathbb I}_2{\mathbb I}_4\big\rbrack \nn \\
&&+2c_2 C_{\scriptscriptstyle{LO}\, 0} {\mathcal G}^2(k)\, \big\lbrace 2{\mathcal G}^2(k) k^2
+ C_{\scriptscriptstyle{LO}\, 0}{\bm Z}^{-1}\lbrack   {\mathcal G}^2(k)\left(2k^2{\mathbb I}_2-{\mathbb J}_2\right)+2k^2{\mathbb I}_4-{\mathbb J}_4\rbrack \nn \\
&&+ C^2_{\scriptscriptstyle{LO}\, 0}{\bm Z}^{-2}\lbrack 2{\mathbb I}_2{\mathbb I}_4 k^2-{\mathbb I}_2{\mathbb J}_4 -{\mathbb J}_2{\mathbb I}_4 \rbrack
\big\rbrace
\label{eq:TNLObare}
\end{eqnarray}
where
\begin{eqnarray}
{\mathbb I}_4 & =& M\int \frac{d^3q}{(2 \pi)^3} \frac{{\mathcal G}^4(q)}{k^2- {q^2}+i\epsilon} \ =\ -\frac{M}{4\pi}{\mathcal G}^4(k)\left(\oneht\mu -\frac{k^2}{2\mu}+i k\right)  \ \ ,\nn \\
  {\mathbb J}_2 & =& \left(\frac{\omega}{2}\right)^{3-d}M\int \frac{d^dq}{(2 \pi)^d} {\mathcal G}^2(q) \ \  \mapup{\rm PDS}\ \  -\frac{M\mu^2}{4\pi}\left(\mu-\omega \right) \ \ , \nn \\
  {\mathbb J}_4 & =& M\int \frac{d^3q}{(2 \pi)^3} {\mathcal G}^4(q) \ =\ \frac{M\mu^3}{8\pi} \ .
\label{eq:divints}
\end{eqnarray}
Here the linearly divergent integral ${\mathbb J}_2$ has been evaluated in dimensional regularization with the PDS scheme~\cite{Kaplan:1998tg,Kaplan:1998we},
and $\omega$ is the renormalization scale\footnote{The linearly divergent integral ${\mathbb J}_2$ in the $\overline{MS}$ scheme can be obtained by setting $\omega = 0$. Similarly, cutoff regularization, as in Eq.~(\ref{eq:In}), is obtained by replacing $\omega$ with $\Lambda$ (for $\Lambda$ large).}.
In terms of renormalized parameters, the amplitude takes the form
\begin{eqnarray}
T_{\NLOss} &=&   C_{\scriptscriptstyle{LO}\, 0} \left( {\mathcal G}^2(k) {\bm Z}^{-1}\right)^2\big\lbrack c_0^R\left(1-\zeta \nu/\mu\right)+ c_2^R \left(3-\zeta \nu/\mu\right)k^2 \big\rbrack \ ,
\label{eq:TNLOrenorm}
\end{eqnarray}
where the renormalized parameters, $c^{(\prime)R}_m$,  are defined as
\begin{eqnarray}
c_2^R  &=&  c_2 \ \ \ , \ \ \ c_0^R  \,=\,  c_0(\omega) + {c_2^R}\left(\frac{\mu+ \zeta \nu}{\mu-\zeta \nu}\right)\big\lbrack \zeta \mu \nu +\omega\left( \mu-\zeta \nu\right)\big\rbrack \ .
\label{eq:renormpars}
\end{eqnarray}
Matching to the expanded ERE of Eq.~(\ref{eq:eftNLOrr2}) gives $c_0^R=0$\footnote{Note that one can also decompose $a=a_{\LOss}+a_{\NLOss}$ in which case this condition follows from $a_{\NLOss}=0$.} and 
\begin{eqnarray}
C_{\scriptscriptstyle{NLO}\,2}\,=\, \frac{M}{4\pi} C^2_{\scriptscriptstyle{LO}\,0}\left(3-\zeta \nu/\mu\right)^{-1}\oneht r_{\NLOss} \ ,
\label{eq:renormparsere}
\end{eqnarray}
with $C^{(\prime )}_{\scriptscriptstyle{NLO}\,m}  \,=\, C_{\scriptscriptstyle{LO}\,0}\cdot c^{(\prime )R}_m$.
This relation gives a subleading enhancement of the same form as the usual pionless EFT~\cite{vanKolck:1998bw,Kaplan:1998tg,Kaplan:1998we} up to the factor in parenthesis,
and results in the scaling
\begin{eqnarray}
c_2^R \ \sim \ \frac{1}{{\mathcal M} \aleph } \ \ \ , \ \ \ C_{\scriptscriptstyle{NLO}\,2} \sim \frac{4\pi}{M {\mathcal M}\aleph^2 } \ .
\label{eq:scaleNLO}
\end{eqnarray}

In similar fashion, the NLO amplitude of Eq.~(\ref{eq:eftNLOrr}) can be obtained via the 
energy-independent residual potential
\begin{eqnarray}
  V_r(p',p)&=& \big\lbrack c_0+ c_2 \left(p^2 + p'^2\right)+c_4 \left(p^4 + p'^4\right)+c'_4 p^2 p'^2+\dots \big\rbrack \nn \\
  &&\times V_{\LOss}\left(p',p\right)\left({\mathcal G}^2(p)+{\mathcal G}^2(p')\right)\ .
\label{eq:VexpEF2c4}
\end{eqnarray}
Working in dimensional regularization with $\overline{\textstyle{MS}}$,
the amplitude takes the form
\begin{eqnarray}
  T_{\NLOss} &=&   C_{\scriptscriptstyle{LO}\, 0} \left( {\mathcal G}^2(k) {\bm Z}^{-1}\right)^2\big\lbrack c_0^R\left(1-\zeta \nu/\mu\right)+ c_2^R \left(3-\zeta \nu/\mu\right)k^2 \nn \\
&&\qquad\qquad\qquad\qquad\qquad\qquad\qquad  + \lbrack 2 c_4^{\prime R} + c_4^R  \left(3- \zeta \nu/\mu\right)\rbrack k^4
  \big\rbrack \ ,
\label{eq:TNLOrenormcasev}
\end{eqnarray}
with the renormalized parameters 
\begin{eqnarray}
  c^{(\prime)R}_4  &=&  c^{(\prime)}_4  \ \ \ , \ \ \
  c_2^R \,=\,  c_2 - \mu \left(\mu+ \zeta \nu\right)\big\lbrack  c_4^R + \left(3- \zeta \nu/\mu\right)^{-1}c_4^{\prime R} \big\rbrack \ , \nn \\
  c_0^R  &=& 
  c_0 - \mu\left(\frac{\mu+ \zeta \nu}{\mu-\zeta \nu}\right)\big\lbrack
  -c_2 \nu \zeta+c_4^{R} \mu^2 (2 \mu+\nu \zeta)+c_4^{\prime R} \mu^2 (\mu+\nu \zeta)
  \big\rbrack \, .
  \label{eq:renormparscasev}
\end{eqnarray}
Matching to the expanded ERE of Eq.~(\ref{eq:eftNLOrr}) now gives $c_0^R=c_2^R=0$ and 
\begin{eqnarray}
C_{\scriptscriptstyle{NLO}\,4}+2C_{\scriptscriptstyle{NLO}\,4}'\left(3-\zeta \nu/\mu\right)^{-1} &=& \frac{M}{4\pi} C^2_{\scriptscriptstyle{LO}\,0} \left(3-\zeta \nu/\mu\right)^{-1} v_2 .
\label{eq:renormparserecasev}
\end{eqnarray}
The coefficients scale as
\begin{eqnarray}
c_4^{(\prime)R} \ \sim \ \frac{1}{{\mathcal M}^3 \aleph } \ \ \ , \ \ \ C^{(\prime)}_{\scriptscriptstyle{NLO}\,4} \sim \frac{4\pi}{M {\mathcal M}^3\aleph^2 } \ .
\label{eq:scaleNLOcasev}
\end{eqnarray}
This differs from the conventional pionless theory counting which has a nominally leading contribution to the
$C^{(\prime )}_{4}$ operators from effective range (squared) effects.

%%%%%%%%%%%%%%%%%%%%%%%%%%%%%%%%%%%%%%%%%%%%%%%%%%%%%%%%%%%%%%%%%%%%%%%%%%%%%%%%%%%%%%%%
\section{EFT description: the NN s-wave phase shifts}

\begin{figure*}
 \centerline{\includegraphics[width=0.98\textwidth]{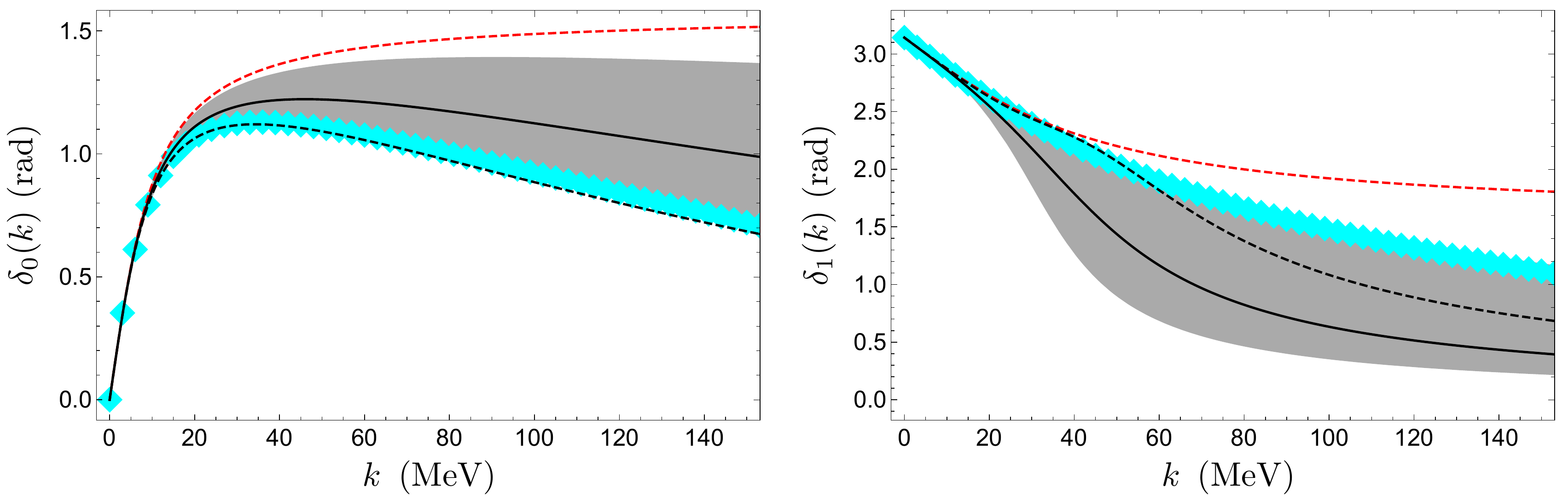}}
 \caption{Singlet (left panel) and triplet (right panel) s-wave NN
   phase shifts with the NLO amplitudes given by
   Eq.~(\ref{eq:eftNLOrr2}) plotted versus center-of-mass momentum
   $k$. The cyan polygons are the Nijmegen phase shift analysis
   (PWA93)~\protect\cite{NNOnline}. The red dashed curve is the LO
   phase shift in the pionless theory (scattering length only). The solid black
   curve is the central value of the LO phase shifts with exact UV/IR symmetry. The gray region represents an error band for the black curve and the dashed black
   curve is the NLO curve.}
\label{fig:LOsingtrip}       % Give a unique label
\end{figure*}

\noindent The phase shifts to NLO in the EFT expansion are
\begin{eqnarray} 
\delta_s(k)&=&  \delta_{\LOss\, (s)}(k)\, +\,  \delta_{\NLOss\, (s)}(k) \ ,
  \label{eq:phaseshifts1}
\end{eqnarray}
with
\begin{eqnarray} 
  \delta_{\LOss\, (s)}(k)&=& -\frac{1}{2} i \ln\left(1 \ - \ i \frac{k M}{2\pi} T_{\LOss\, (s)}(k) \right) \ , \\
  \delta_{\NLOss\, (s)}(k)&=& -\frac{k M}{4\pi} T_{\NLOss\, (s)}(k) \left(1 \ - \ i \frac{k M}{2\pi}  T_{\LOss\, (s)}(k) \right)^{-1}  \ .
  \label{eq:phaseshifts2}
\end{eqnarray}
Recall from  section~\ref{sec:uvir} that in s-wave NN scattering, the UV/IR symmetry of the full $S$-matrix
requires range corrections that are correlated with the scattering lengths and treated exactly.
The physical effective ranges can therefore be expressed as $r_s=r_{\LOss\,(s)}+r_{\NLOss\,(s)}$ with
\begin{eqnarray} 
r_{\LOss\,(0)}&=&2\lambda a_1 \ \ \ ,\ \ \ r_{\LOss\,(1)}\,=\,-2\lambda a_0\ .
  \label{eq:confmoebiusiso4}
\end{eqnarray}
The singlet and triplet phase shifts with the NLO amplitude given by Eq.~(\ref{eq:eftNLOrr2}) are plotted in Fig.~(\ref{fig:LOsingtrip}).
At LO, there are three parameters given by the two s-wave scattering lengths and $\lambda$, which is fixed to
$\lambda=0.14\pm 0.11$, the range of values that exactly encompasses fits of $\lambda$ to each s-wave channel independently. This spread in $\lambda$
corresponds to the shaded gray region of the figures and is a conservative estimate of the LO uncertainty. The NLO curve has been generated
by tuning the $r_{\NLOss\,(s)}$ to give the ``physical'' effective ranges. A band on the NLO curves can easily be set by folding in the nominally
NNLO effect.

\begin{figure*}
 \centerline{\includegraphics[width=0.98\textwidth]{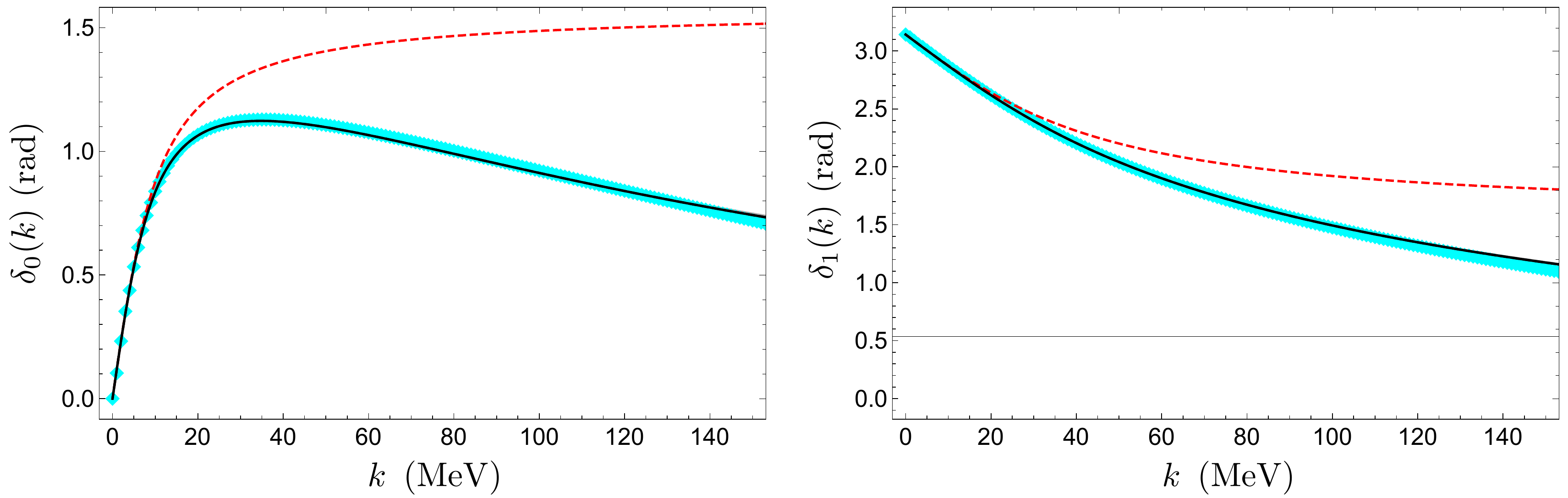}}
 \caption{
   Same as in Fig.~(\ref{fig:LOsingtrip}) but with the effective ranges treated exactly in each channel and the NLO amplitudes given by Eq.~(\ref{eq:eftNLOrr}).
   In both channels, the NLO curve overlaps with the central value of the LO phase shift.}
\label{fig:LOsingtripcasev}       % Give a unique label
\end{figure*}

The singlet and triplet phase shifts with the NLO amplitude given by Eq.~(\ref{eq:eftNLOrr}) are plotted in Fig.~(\ref{fig:LOsingtripcasev}). As this treats both effective ranges exactly it provides an extremely accurate fit to the phase shifts due to the smallness of the shape parameters.

%%%%%%%%%%%%%%%%%%%%%%%%%%%%%%%%%%%%%%%%%%%%%%%%%%%%%%%%%%%%%%%%%%%%%%%%%%%%%%%%%%%%%%%%
\section{Conclusion}

\noindent The s-wave NN $S$-matrix obtained from the ERE with
scattering length and effective range terms only, has interesting
UV/IR symmetries which are special inversions of the momenta. These
symmetries set the region of applicability of the EFT
descriptions. For instance, while it is common to view the EFT of
large scattering lengths as an expansion about the unitary fixed point
of the RG, it is, strictly speaking, an expansion about the fixed
point of a momentum inversion symmetry.  The UV/IR symmetries, while
not symmetries of the EFT action or of the scattering amplitude, are
present in the interaction. For instance, in the EFT of large
scattering lengths, the UV/IR symmetry is manifest in the RG flow of
the contact operator as an inversion symmetry of the RG scale which
interchanges the trivial and unitary RG fixed points and leaves the
beta function invariant~\cite{Beane:2021xrk}. When the effective range is also treated at LO in the EFT, the $S$-matrix has a (distinct) UV/IR symmetry which
effectively determines the LO potential, and constrains the form
of the perturbative NLO corrections.

There are many avenues to pursue with this new EFT. For instance, the softened asymptotic behavior of
the LO potential may resolve the issues of renormalization that arise when range corrections are
added to the integral equations that describe the three-nucleon system at very low energies. In addition, given the improved
convergence of LO in the EFT up to momenta beyond the range of validity of the pionless EFT, the perturbative pion paradigm may be worth revisiting in this scheme.
It may also be the case that the UV/IR
symmetries have interesting consequences for systems of many nucleons near
unitarity.

\begin{acknowledgements}
We would like to thank Daniel R.~Phillips for a careful reading of
the manuscript and many useful comments and suggestions. In addition
we are grateful to Bira van Kolck and Ulf G.~Mei\ss ner for interesting comments and for pointing out important missing
references.
This work was supported by the U.~S.~Department of Energy
grants {\bf DE-FG02-97ER-41014} (UW Nuclear Theory, NT@UW-21-16) and {\bf DE-SC0020970}
(InQubator for Quantum Simulation, IQuS@UW-21-017).
\end{acknowledgements}

\bibliographystyle{spphys}       % APS-like style for physics
\bibliography{bibi}   % name your BibTeX data base

%\begin{thebibliography}{}
% etc
%\end{thebibliography}

\end{document}